# Engineering the Future of R&D: The Case for AI-Driven, Integrated Biotechnology Ecosystems

Alex Zhavoronkov[1*], Chuen Yan Leung[2]

[1]Insilico Medicine, Cambridge, MA, USA
[2]Silver Dart Capital, Hong Kong SAR, China

Correspondence: alex@insilico.com

# Abstract

The escalating cost, extended timelines, and low success rates in pharmaceutical research demand a fundamental rethinking of biotechnology R&D infrastructure. This chapter introduces the concept of the AI-Integrated Biotechnology Hub, a purpose-built research ecosystem uniting residential, commercial, clinical, and research facilities under a central, AI-driven operating system. The hub functions as a multi-sided platform that unites health data collection, smart living environments, and federated learning to enable secure, privacy-preserving biomedical research. By integrating real estate, biotechnology facilities, research hospitals, and community services, the model maximizes data utility, accelerates drug discovery, and enhances resident well-being. Transparency, accountability, and ethical stewardship are critical pillars of governance, enacted through dynamic consent, data trusts, and multi-stakeholder oversight. Scalable across urban and vertical architectures, this paradigm offers a viable, sustainable pathway toward improving healthspan, fostering innovation, and reshaping the economics of global drug development.

# Introduction

The economics of pharmaceutical research and development (R&D) are unsustainable. The cost to bring a single new drug to market has grown exponentially, from an estimated $1.8 billion in 2010 (Paul et al. 2010) to over $6.1 billion by 2020, with timelines stretching from 10 to 20 years. Compounding this, the clinical success rate remains stubbornly low, with only 8% of drugs entering clinical trials ultimately gaining regulatory approval (Maharao et al. 2020). This high-risk, high-cost paradigm deters investment, stifles innovation, and hinders the translation of scientific breakthroughs into tangible health benefits. Cyclic industry trends drain interest and investment in promising companies and products at off-peak times (Gehrig and Stenbacka 2003; Hine and Griffiths 2004) and hinder maintenance of a top talent pool in light of such unpredictable outcomes and unstable investment (Cao 2007). To reverse this trend, a paradigm shift in the foundational infrastructure of drug discovery is required.

Not only does the biomedical research industry require a shift in strategy to continue to sustainably grow and withstand market fluctuations and pressures, but the output of research is intimately tied to society-wide productivity. Economic models have shown that the rate of biomedical innovation and, more importantly, the extending of healthy, productive lifespan are directly linked to economic growth (Zhavoronkov and Litovchenko 2013). Rapid adoption of biomedical discoveries and investment in longevity research contribute to this growth (Goldman et al. 2013; Zhavoronkov and Litovchenko 2013).

Here, we propose a new model for biotechnology R&D: a fully integrated, AI-driven ecosystem that converges residential, commercial, clinical, and research infrastructure. This "Integrated Hub" is conceived as a purpose-built environment designed to systematically eliminate the friction points that plague the current R&D landscape. By leveraging real estate assets to create a physically and digitally unified community, this model aims to de-risk investment, accelerate discovery, and maximize the value of health data, ultimately creating a more efficient and productive innovation engine.

# The Limits of Current Innovation Models

The transformation of the traditionally low-yield but stable investment in real estate (RE) into higher-yield biotechnology infrastructure is the perfect opportunity to reprioritize historically unproductive assets into transformative biopharmaceutical and healthcare innovation. The prevailing models for fostering biotech innovation rely on organic clusters and specialized real estate investment trusts (REITs) and have historically been successful but are now reaching the limits of their efficacy. These models are rooted in the economic theory of agglomeration, where the co-location of firms, talent, and capital generates positive externalities through the sharing of resources, better labor market matching, and knowledge spillovers (Duranton and Kerr 2015; Bolter and Robey 2020).

Kendall Square in Cambridge, Massachusetts, stands as the archetypal organic innovation cluster. Its dynamism is an emergent property, arising from the dense co-location of world-class universities (MIT, Harvard), medical centers, hundreds of biotech firms, and a deep venture capital pool (Bevilacqua and Pizzimenti 2019; Buderi 2022). However, its success has created significant negative externalities, including prohibitive housing costs, a scarcity of lab space, and severe congestion, which now threaten to impede further growth (Ledford 2015). Governance is also fragmented, with multiple stakeholders like university committees and local associations making coordinated solutions difficult (Bevilacqua and Pizzimenti 2019).

A more deliberately curated approach is offered by specialized life science REITs, most notably Alexandria Real Estate Equities (ARE). Far from being a passive landlord, ARE actively develops "Megacampus" ecosystems within top innovation clusters (Philippidis 2024). ARE’s model extends beyond real estate to include a venture capital arm that invests in its tenants and a thought leadership platform to foster collaboration, creating a powerful, synergistic business model (Philippidis 2024). Yet, while ARE provides premier infrastructure and capital, it does not, and cannot, address the core inefficiencies within the R&D process itself. Data remains siloed within tenant companies, and there is no mechanism to integrate the rich longitudinal data from the community's daily life into the research engine.

The Integrated Hub model is designed to overcome the limitations of both organic and curated clusters by introducing a new layer of deliberate, data-driven integration.

# Maximizing the value of health data in an AI-driven biotechnology research ecosystem

Each modality of health data (e.g., images, videos, voice, blood tests) has its own predictive strength, but its value declines over time, much like a half-life (Mamoshina et al. 2018). Combining data types collected over time can enhance predictive accuracy by creating a synergistic effect. Even better, combining continuous data collection overcomes limitations imposed by fragmented or sporadic collection. An environment facilitating frequent data collection will maximize the synergistic benefit patients and providers get for any mode of health data available.

Machine learning-driven AI systems enabling the integration of these disparate and high-dimensional data streams (Baltrusaitis et al. 2019). Deep neural networks (DNNs) trained on multi-modal data (photographs, videos, lab tests, "omics," etc.) can identify biologically relevant features related to aging and disease. These networks can be used to discover key features for disease prediction, build association networks, and establish causal links between variables (Mamoshina et al. 2018).

How to fully leverage digital health data and distribute monetized value back to the various stakeholders– patients, medical providers, investors– remains an unresolved question (Blumenthal 2017). As such, formal models for valuing and pricing human health data are in development, and these models would necessitate a data marketplace where buyers use the data to research and create diagnostic and therapeutic products (Zhavoronkov and Church 2019). Indeed, attempts have been made to put human health data on the blockchain for the purposes of research and monetization (Zhavoronkov and Church 2019). An integrated health monitoring and data distribution ecosystem would streamline the internalization and externalization of the data, all controlled and tracked by a central processing hub, to return maximized value to both the patient and biotechnology partners for personalized and generalized health solutions.

A centralized hub model of the biotechnology ecosystem aims to boost connectivity, efficiency, and communication within and across stakeholders and industry players by integrating every level of personalized medicine, together leading to synergistic value growth. Moreover, currently implemented AI-driven personalized healthcare systems often fail to identify and use the most appropriate data for intended purposes, resulting in limited value from healthcare data assets.

The process of data integration, cleansing, interpretation, and aggregation is often duplicated by researchers, leading to inefficiencies and repeated challenges (Ozaydin et al. 2020).

A shared network hub model alleviates these inefficiencies with its single-operator design and tailor-made AI operations constantly refreshing the learning algorithms to return the most informative, actionable insights at every level (Figure 1). At the most granular, personal level of the model, smart technology and smart city infrastructure communicate lifestyle and behavioral prescriptions directly to the people. While not novel concepts or technology, applying this framework to healthcare with smart biotechnology opens the door to immediate health value for users and research organizations. At the next layer, smart biomedicine capitalizes on high-dimensional, longitudinal biometric data for local analysis and delivery of optimized healthcare to individuals, identifying risk biomarkers before disease onset when preventative treatments or lifestyle measures are most effective. Collecting biometric and lifestyle data from entire communities leads to insights for the smart biotechnology industry, leveraging big data computation for biomedical discovery and the development of future medical interventions. The ultimate peak in health data analytics in this framework is smart longevity systems, improving healthspan with the solutions realized from every lower layer of insight. These solutions can only be practically learned and implemented through a centralized hub design and the reduced barriers to data collection, analysis, sharing, and distribution entailed.

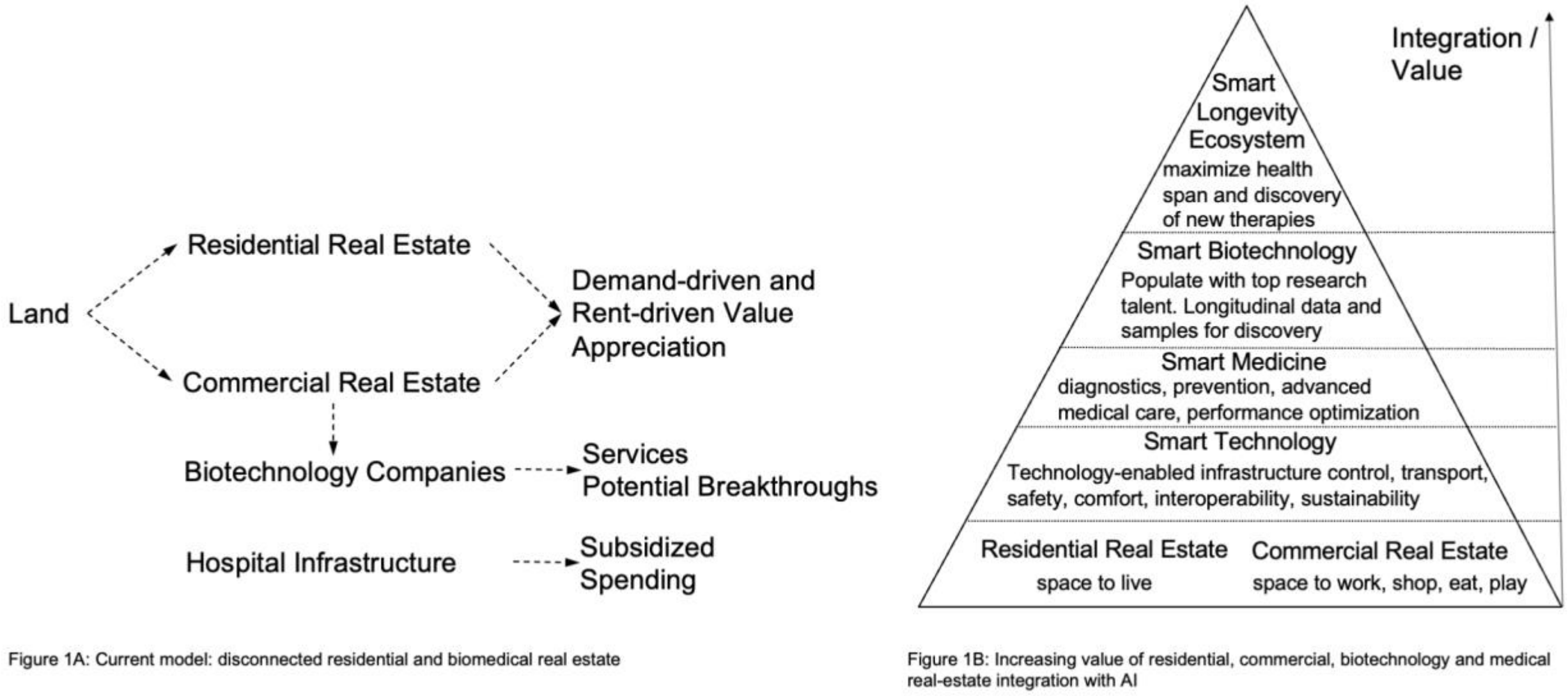

Figure 1A: Current model: disconnected residential and biomedical real estate

Figure 1B: Increasing value of residential, commercial, biotechnology and medical real-estate integration with AI

Figure 1. From traditional real estate supply-demand driven property appreciation model to appreciation with significant research contribution and appreciation with potentially uncapped potential

# The New Paradigm: The Integrated Biotechnology Hub

The Integrated Hub is a purpose-built, self-contained ecosystem founded on five key infrastructure elements: (i) advanced residential real estate, (ii) commercial and retail amenities, (iii) biotechnology research facilities, (iv) a research hospital, and (v) a central, AI-based operating system that unifies the entire community.

Conceptually, the hub functions as a Multi-Sided Platform (MSP), a business model that creates value by connecting distinct user groups: researchers (who are also residents), biotech firms, the hospital, and commercial tenants (Teece et al. 2022). This framework is common in the digital economy but novel in its application to physical R&D infrastructure. As an MSP, the hub's value is driven by network effects: more residents generate more high-quality longitudinal data, which makes the platform more valuable to biotech firms, which in turn can fund better amenities and research tools, attracting more top talent (Robinson and Stuart 2006).

This model presents a fundamental strategic trade-off. The "single operator" structure enables vertical integration for maximum efficiency and coordination. However, to foster innovation, it must also function as an open platform that motivates the creativity of its participants (Andrei Hagiu 2015).

## Advanced residential real estate in the biotechnology hub

A central factor in the success of the biotechnology hub is the continued health and happiness of the residents– that is, the research staff whose work in the hub directly contributes to sustained and fruitful biomedical innovation. As full-time residents of the ecosystem, researchers themselves would live in the facility's smart homes. These homes provide personalized feedback on biometric and lifestyle data, identifying disease-associated deviations and providing lifestyle coaching, while also feeding information back into the central AI for continuous refinement (Figure 2).

Today's Internet of Things (IoT) technology has hinted at the potential of personalized healthcare by enabling personalized, in-home monitoring systems utilizing wireless health sensors to continuously track biometric parameters like pulse rate, blood pressure, and body temperature (Kelly et al. 2020). IoT architectures integrate smart home devices with healthcare services, allowing for data collection, processing, and sharing with medical professionals (Datta et al. 2015). This approach offers numerous benefits, including improved safety, quality of life, and reduced hospitalization costs (Philip et al. 2021). IoT-based health monitoring in smart homes has the potential to alleviate strain on traditional healthcare systems by facilitating remote treatment and recovery (Linkous et al. 2019). Recent advances in non-contact monitoring systems solely using video data have opened the possibility to collect biometric data without invasive or disruptive procedures (Siam et al. 2020; Sasaki et al. 2024; Pham et al. 2024).

Gaps in traditional healthcare delivery can be overcome with the integration of AI and machine learning tools into the analysis of each individual's lifestyle and biometric data, offering a more proactive and personalized approach to health management (Jagadeeswari et al. 2018). Advances in IoT, networking, and cloud computing infrastructure have enabled and streamlined sustainable and secure collection and distribution of the vast amounts of data collected by biosensors (Zahid et al. 2022; Khan et al. 2022). Such high-dimensional data collection can be fed into AI systems for automated analysis and detection of anomalous data points. Connection to the ecosystem's global AI operating system will allow seamless integration with medical providers to notify resident's healthcare team of any actionable or noteworthy measurements that merit further, in-person follow-up (Vegesna et al. 2017; Sivani and Mishra 2022).

Additionally, residence facilities require high-level security infrastructure to safeguard the residents' well-being, the integrity of the ecosystem, and the secure operations of the on-site research facilities. Biometric-based identification is a standard component of many of today's security protocols across industries, and the next logical step is the integration of healthcare monitoring into the same systems. Algorithms for devising biometric-based encryption keys to secure biosensor data suggest digital security can be easily integrated into such systems (Pirbhulal et al. 2018a, b), but further leveraging every security validation and as opportunity to assess a resident's health is paramount for gathering as much actionable data for predicting health outcomes. For example, assessing cardiovascular health via retinal blood vessel morphology can be accomplished in the background while scanning for iris patterns for secure admission.

So far, the adoption rate of smart homes has additionally remained low due to technological fragmentation within the smart home ecosystem. The presence of multiple protocols and standards can confuse consumers, making it challenging to choose compatible products (Phan and Kim 2020). Current IoT systems have been limited to the retrofitting of existing architecture or wearable devices, but creating living spaces with health monitoring systems integrated from the initial design and with a single controller entity will alleviate these hurdles and allow for more seamless, diverse, and non-intrusive personalized healthcare capabilities.

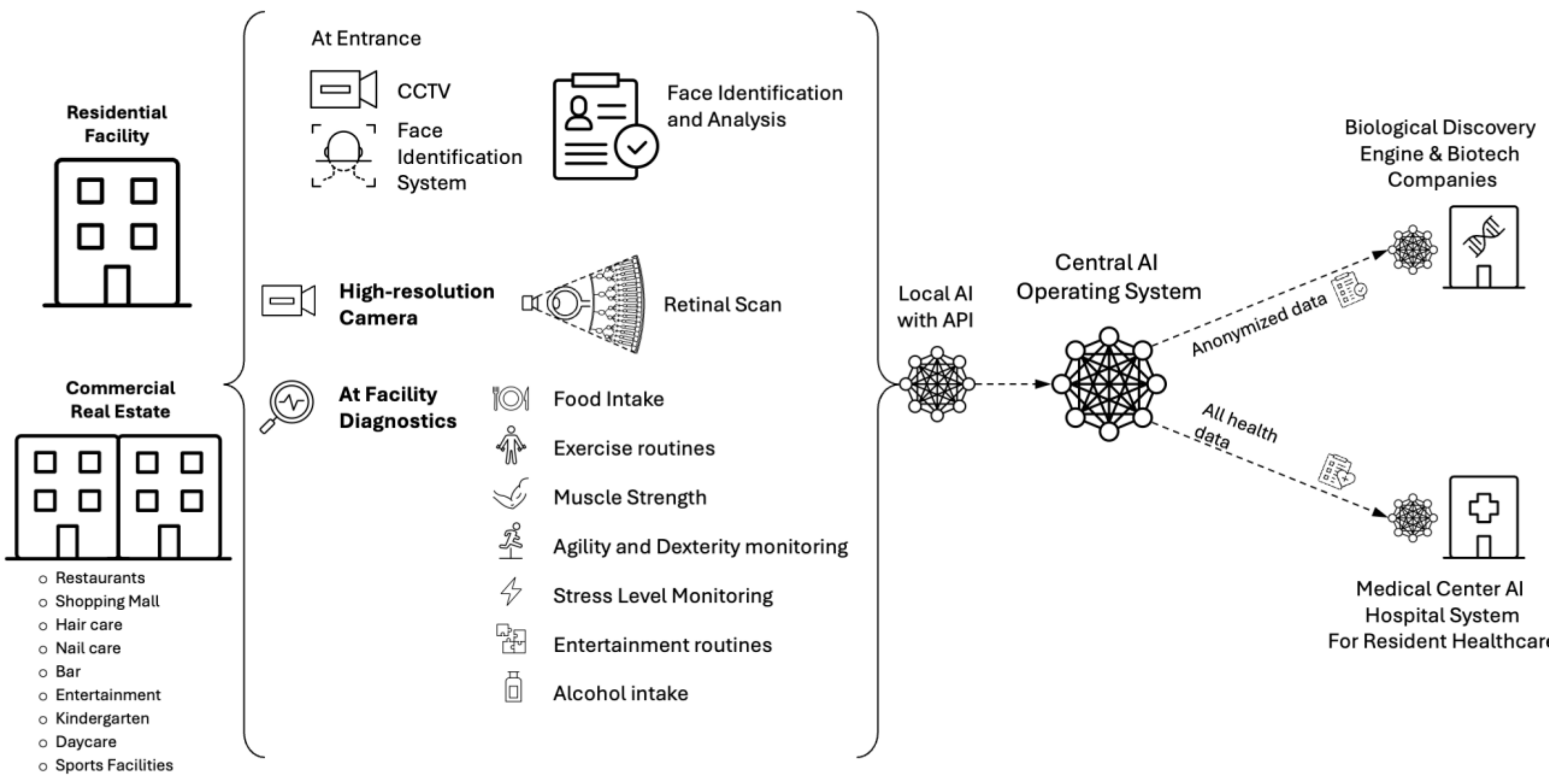

Figure 2. Multi-layer generative AI-driven continuous health monitoring and intervention systems providing valuable data to biotechnology research organizations and medical research centers. This ecosystem provides unprecedented health benefits to the residents and workers of the community while increasing the probability of biomedical breakthroughs arising from longitudinal data and close cooperation between biotechnology companies and medical institutions.

## Commercial and retail real estate in the biotechnology hub

Attracting and retaining the world’s leading biotechnology researchers will partly depend on an integrated approach that incorporates a balance of commercial, retail, and lifestyle amenities to promote health and satisfaction.

Supermarkets and organic food markets are essential to ensure workers have convenient access to fresh, healthy food. The diverse workforce of a biotechnology hub that reflects the diversity of the top talent pool would benefit from an array of health-conscious food and dining options. Convenience stores, pharmacies, and personal care shops should be easily accessible to meet the everyday needs of residents, while specialty stores should offer fitness gear, supplements, and wellness products cater to fully outfit residents with the products needed to keep their lifestyles consistently healthy. Bookstores, music shops, or electronics stores can provide leisure and recreation options, enhancing the overall quality of life, and hosting temporary or rotating pop-up stores can keep the retail scene fresh and dynamic, giving local entrepreneurs and startups a platform to reach hub residents. Providing healthy food options (Cena and Calder 2020) and engaging in leisure activities (Fancourt et al. 2021) are proactive, effective strategies for preventing disease and positively influence health.

The availability of these commercial offerings like shopping malls, restaurants, grocery stores, leisure, exercise, and entertainment facilities will not only present residents with such facilities, but it also serves the goal of reducing the need to leave the facility. Consistent monitoring of food intake and physical activity is the best way to ensure that accurate health information is dispensed to the residents and no anomalous risk biomarkers are missed.

## Biotechnology infrastructure

The ultimate goal of the biotechnology hub facility is to maximize healthcare innovation for the betterment of the world's population while providing its own residents with optimal living and working conditions. Central to this objective is providing the resident research companies with a network of resources, policies, and services that support the industry's unique needs.

High-quality, specialized labs with advanced equipment for DNA sequencing, polynucleotide synthesis, cell culture, robotic systems to automate and expedite research, and dry space for bioinformatics systems. To facilitate end-to-end pharmaceutical development, pilot and commercial-scale manufacturing facilities would enable small-batch production and scaling up capabilities critical for translating research into products. This includes Good Manufacturing Practice (GMP)-certified labs and manufacturing or bio-production facilities, essential for producing biologics, pharmaceuticals, and other biotech products.

A successful biotechnology hub must be physically located close to world-class research universities and medical schools. These institutions provide a continuous pipeline of talent, research collaboration opportunities, and technology transfer partnerships. Public-private partnerships, joint research projects, and technology licensing agreements between academia and industry would be critical for translating academic research into marketable biotechnology solutions.

Numerous support industries, such as funding, marketing, legal, and other service providers integral to a vibrant biotechnology industry are crucial. Access to a strong venture capital ecosystem would help ease the burdens of long development timelines and the patient capital necessary to sustain research and commercialization efforts. Biotechnology companies additionally rely on a highly specialized supply chain for raw materials, lab equipment, and chemicals to ensure smooth and reliable operation for uninterrupted research and production. Access to legal services specializing in patent law, contract negotiations, and regulatory compliance specific to the biotech industry would help navigate the legal and regulatory systems. Many biotechnology products and processes require cold chain logistics for storage and transportation, making access to specialized cold chain systems, including secure underground facilities for long-term cryostorage of cells, tissues, organs, and embryos for research and personal medical usage integral to cutting-edge R&D and the people who make up the industry.

## Research hospital-based healthcare

An on-site research hospital would serve the dual purpose of providing residents and visitors with advanced healthcare and leveraging the massive amounts of biometric data for conducting research studies. A central facility to collect, store, and analyze enormously longitudinal patient data, patient tissue samples, and lifestyle monitoring data would be uniquely positioned to conduct comprehensive and innovative research studies. A wide array of healthcare providers and infrastructure falling under one umbrella organization would advance the vision of today's patient-centered shared electronic health records (SEHRs), which can increase the quality of care while reducing costs, and fully in-house healthcare analytics would break down the barriers imposed by cross-institutional data exchange that has so far faced challenges to standardized, distributed provision of medical data (Schabetsberger et al. 2010).

High-throughput biobanking capabilities would enable tissue samples, blood, DNA, and other biological specimens collected from patients to be stored and processed immediately and on-site. Each sample's link to an individual patient's longitudinally collected data would allow researchers to connect genetic, molecular, or lifestyle information with patient outcomes or demographic information. Going far beyond today's electronic health record (EHR) systems, such a system capturing biometric data not only during patient visits but also non-invasively and unobtrusively via smart home monitoring can be enriched by integrating varied clinical data sources like lab results, sequencing data, imaging, and treatment histories, all being automatically fed into and processed together in a single AI-powered personalized medical database in real-time. AI and machine learning algorithms can analyze these integrated datasets to identify patterns and predict disease risk, onset, or progression to support personalized early diagnosis and preventative care strategies.

The centralized nature of the healthcare system would benefit from standardized data formats, ensuring that all data types (biometrics, tissue samples, lifestyle data) are collected and stored to allow for seamless integration, analysis, and access by both medical providers and researchers. The interoperability of such a system ensures that data from different sources (EHRs, wearables, biobank metadata) can be contributed and accessed by APIs to facilitate real-time synchronization between different databases and research platforms.

Hospital researchers can leverage big data analytics to conduct large-scale population health studies by analyzing biometric, genetic, and lifestyle data across thousands of patients. This approach can identify public health trends or the efficacy of interventions across different demographics.

# The Digital Architecture: A Privacy-First, Multi-Agent AI Operating System

At the heart of the hub is a three-tiered AI operating system designed to coordinate research and protect privacy (Figure 3). We envision this as a Multi-Agent System (MAS), a distributed network of autonomous AI agents collaborating to solve complex problems that are beyond the scope of a single agent (Skobelev 2011; Skupin and Metzger 2012; Nweke et al. 2025).

**Personal AI Agents:** These agents reside on users' personal devices, acting as digital stewards. They manage individual privacy preferences and permissions, interacting with the broader system on the user's behalf and maximizing health benefits by providing timely diagnostics and preventative health insights or recommendations.

**Residential and Commercial Hub AI Agents:** These are specialized AI systems operating at the level of facilities– companies, commercial centers, and buildings. A "Residential Agent" could optimize building energy and resource use, a "Hospital Agent" could manage patient flow and clinical trial recruitment, and a "Lab Agent" could orchestrate robotic automation for high-throughput experiments.

**Community-scale Central AI Reasoning and Operating System:** The central AI Operating System would be utilizing the latest multimodal foundational models for reasoning and central orchestration of multiple domain-specific models, agent creation, deployment, and supervision. Central OS planning ensures that the global objectives of the community, such as maximizing research productivity while ensuring resident well-being, align with short- and long-term resource allocation and facilitates coordination among the various deployed agents.

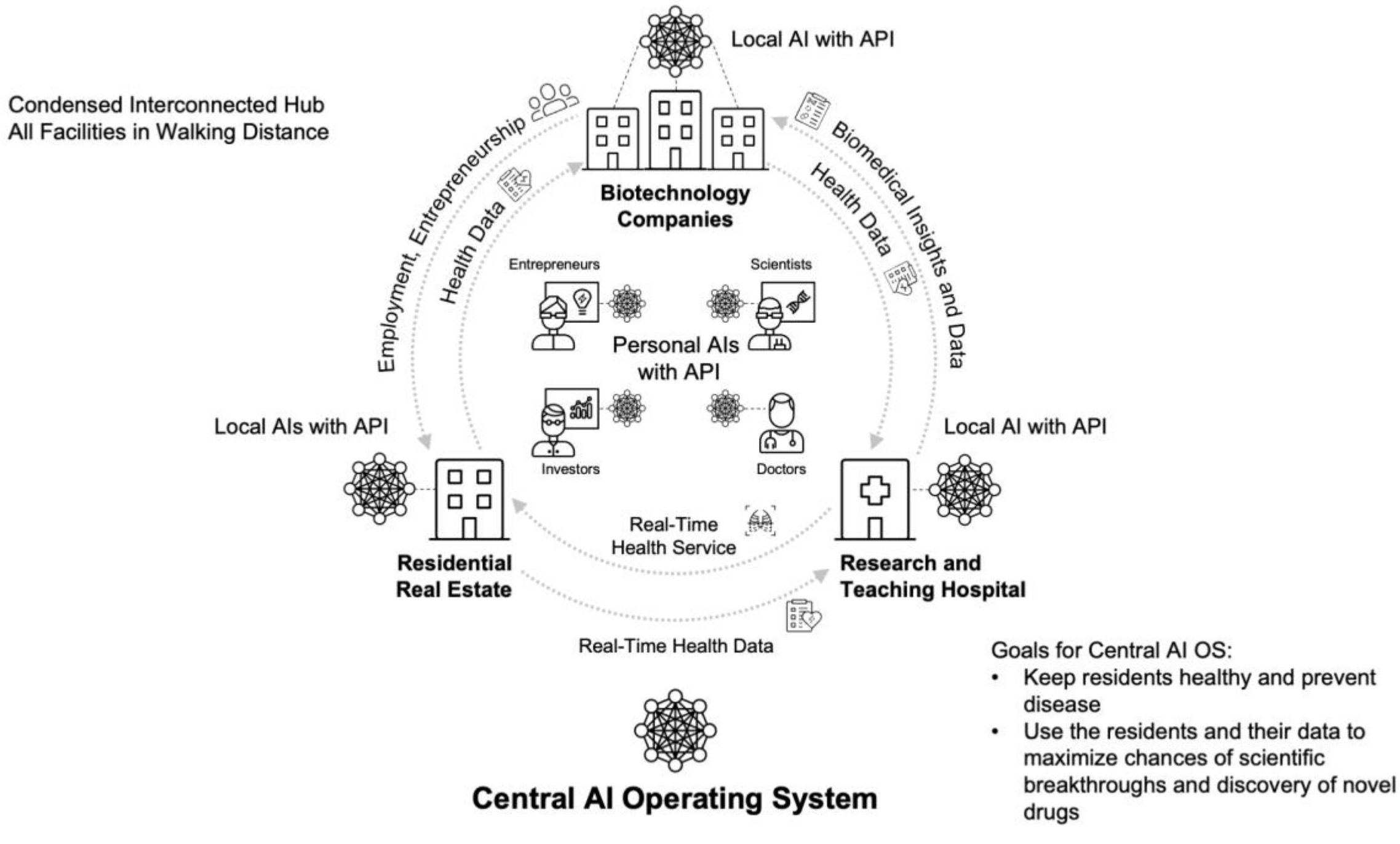

Figure 3. Toward the fully-integrated maximally-efficient frictionless residential-research-biomedical ecosystem in a three-tiered AI operating system: combining real-time data collection and health monitoring with biomedical research and clinical services in a federated learning-based smart ecosystem.

# Ethical Considerations for Ecosystem Stewardship

A critical flaw in any model that centralizes health data is the immense privacy and security risk. Data security must be a priority in the design of the research hub architecture, as the massive amounts of personal data collected on residents and visitors are a prime target for bad actors to destroy or steal.

To address this, the hub's architecture must be built on a foundation of Federated Learning (FL). FL is a machine learning paradigm where the AI model is trained on *decentralized* data sources without the raw data ever leaving its secure environment. In the hub, the hospital, individual biotech firms, and even residents' personal devices would serve as nodes in a federated network. A global model, for instance, one predicting disease risk, could be trained collaboratively across all nodes without any sensitive patient or proprietary corporate data being exposed (Andreux et al. 2020; Oldenhof et al. 2023; Choudhury et al. 2025). This privacy-preserving-by-design approach transforms the model's greatest potential weakness into a state-of-the-art strength, enabling collaboration between competing entities while maintaining data sovereignty (Loftus et al. 2022).

Additionally, recent developments in health technology security have advanced the ability to detect attacks on IoT systems, for example, by machine learning models profiling IoT data traffic patterns (Khan and Alkhathami 2024). Such proactive monitoring measures must complement the more traditional preventative security measures, such as encryption, physically and digitally secure databases, and appropriately air-gapped internal networks.

A single authority overseeing a combined residential, industrial, and commercial facility raises important concerns about privacy, data ownership, and governance. While AI-driven health monitoring can provide timely and accurate information for managing chronic conditions like Parkinson's disease, it also raises questions about evolving illness identities and care relationships (Ho et al. 2024). Although many individuals are willing to share their data for research purposes, there is a lack of awareness about data access and use by companies (Nebeker et al. 2020). The digital health ecosystem presents challenges in selecting, testing, and implementing technologies ethically (Nebeker et al. 2019). To address these issues, stakeholders must prioritize transparency, informed consent, and respect for persons in digital health research (Nebeker et al. 2019, 2020).

Ensuring that personal health data are transparently and appropriately collected, organized, and used is important for residents' trust in the system and continued partnership and participation. Many organizations find it difficult to establish robust data governance infrastructures to meet the increasing demands of advanced data uses, which impedes the sharing of data across different entities (Tumulak et al. 2024). Moreover, issues related to data distribution and accessibility continue to pose problems, despite substantial investments aimed at making healthcare data available to clinicians and researchers (Newman et al. 2014). Importantly, any governance and ethical structures implemented need to consider the local or regional societal priorities, beliefs, and legal frameworks on the topic, as different populations have different levels of trust and expectations of healthcare AI systems (Char 2022).

# Governance for a Trustworthy Ecosystem

With an unprecedented range of AI agents and real-world stakeholders able to access residents' data, a concerted effort needs to also be made to receive truly informed consent before residents' biometric data, tissue samples, and lifestyle information are collected and used for research and healthcare purposes that go far beyond the uses and limitations of their data today. Acknowledging the inherent risks is the first step toward building a trustworthy and socially sustainable ecosystem. The governance framework must move beyond technocratic efficiency to embrace transparency, accountability, and equity (Hodson et al. 2023).

**Hybrid Governance:** The "single operator" model must be balanced by a hybrid governance structure that includes a multi-stakeholder oversight board. This board should be composed of elected resident representatives, members from tenant companies, the hospital's ethics committee, and liaisons from local government (Lee et al. 2023) . This ensures that the operator remains accountable to the community it serves, preventing the emergence of a "post-political" system where decisions are made without public oversight (Ziosi et al. 2022).

**Dynamic Consent and Data Stewardship:** The continuous, longitudinal nature of data collection in the hub renders traditional, one-time "broad consent" ethically insufficient and practically obsolete (Michael Froomkin 2018). In a Big Data context, where future uses of data are unknowable, a more robust framework is required (Michael Froomkin 2018). We propose a move towards participatory models of data stewardship:

> Dynamic Consent: A digital interface that allows residents to granularly manage permissions for their data in real-time, consenting to new research projects or changing their preferences as they see fit (Muller et al. 2023). This transforms consent from a static, one-time event into an ongoing, engaged dialogue.
>
> Data Trusts: A legal structure where residents collectively place their data rights into a trust. An independent trustee is then legally bound by fiduciary duty to manage data access and use in the best interests of the beneficiaries (the residents) (O'Hara 2019; Iqbal et al. 2025). This model provides a robust legal and ethical framework for data stewardship, ensuring accountability and alignment with community values.

# Pilot Projects and Future Directions

## Fully-integrated Wide-area Longevity Biotechnology Hub Concept

The Integrated Hub model is not merely a theoretical construct. Elements of this vision are beginning to take shape in ambitious projects around the world. For instance, large-scale infrastructure projects like TODTOWN in Shanghai's Pudong district, a massive Transit-Oriented Development integrating commercial, residential, and public space, provide a blueprint for the physical infrastructure of such a hub.

Perhaps the most ambitious real-world test of this concept is NEOM in Saudi Arabia, a region being built from the ground up as a "cognitive city" powered by AI and data (Alam et al. 2021). While facing significant challenges in attracting international residents willing to pay premium prices for an untested concept in a remote location (Cafiero 2024), NEOM's strategy offers a compelling solution. It aims to create a unique value proposition centered on health and longevity (Farag 2019). By building an integrated healthcare system with the explicit goal of proactive, predictive, and personalized care, NEOM intends to become a global destination for wellness and longevity research (Alam et al. 2021; Bhaska 2023). The integration of AI with the objective to ensure and extend the healthspan of its residents may not only increase its appeal but also help make history by creating the first city designed for human longevity.

In dense urban environments, the model can be adapted into a vertical architecture, as illustrated by the conceptual "DNA Towers" designed for Hong Kong, which stack residential, commercial, and research facilities into a single, iconic structure. Similarly, a mixed-use development underway in Pudong, Shanghai, China, is set to host IoT-enabled residences, top-of-the-line healthcare facilities, commercial districts, dedicated biotechnology R&D spaces, and on-site hospital facility (Figure 4). Lessons learned from the design, construction, implementation, and research output of this site, preliminarily named “TopTown,” will inform future research hub design.

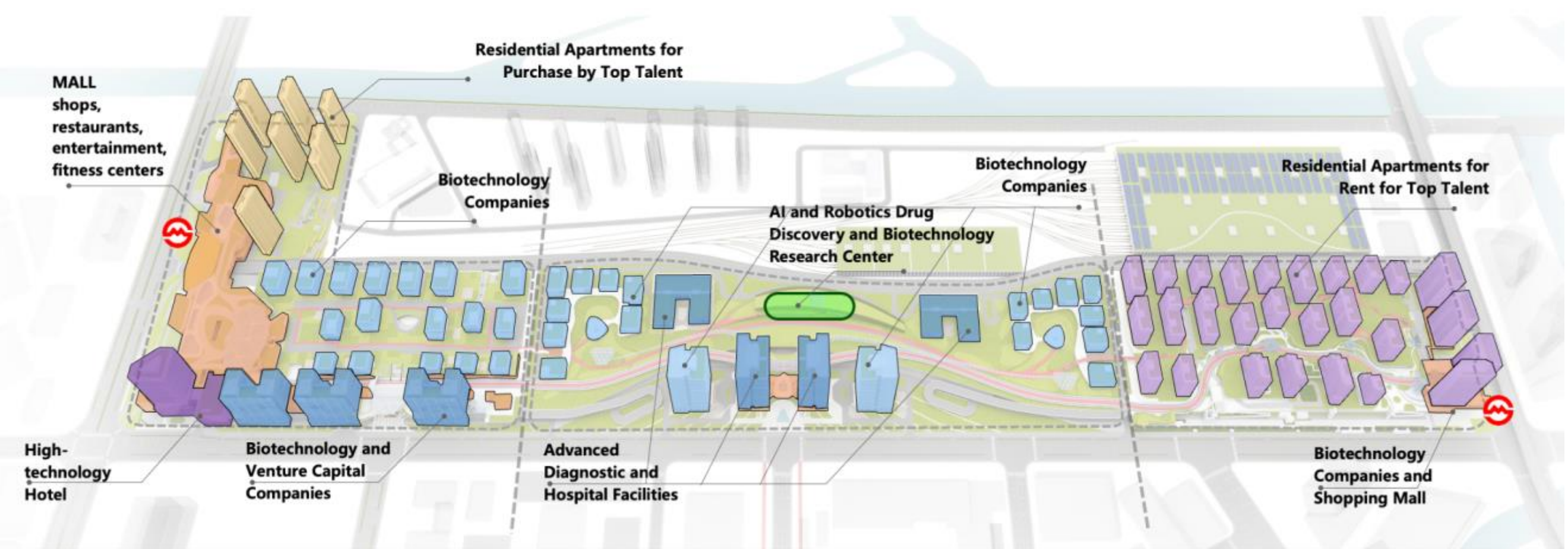

Figure 4: Conceptual overview of a fully-integrated residential, commercial, biotechnology and biomedical infrastructure based on Shanghai Pudong Development Group's TopTown Project

## The Biotechnology Hub Tower

A focus on vertical architecture to house the hub allows the stacking of industrial, office, and residential spaces in a single structure, letting businesses operate more efficiently, reducing transportation needs and consolidating supply chains, particularly in places where both real estate costs and population density are the highest (Kodmany and Ali 2012) (Figure 5). This approach to mixed-use development has already gained traction in various cities around the world as a way to boost the land value of inefficiently planned urban centers, including the recently constructed Shenzhen Bay Innovation and Technology Centre (Yang and Garzik 2022; Cheshmehzangi and Tang 2024).

Realizing this vision requires unprecedented collaboration between real estate developers, technology companies, academic institutions, healthcare providers, and government. However, by fundamentally rethinking the physical and digital infrastructure of R&D, the Integrated Biotechnology Hub offers a viable, if ambitious, path toward a more efficient, equitable, and sustainable future for pharmaceutical innovation.

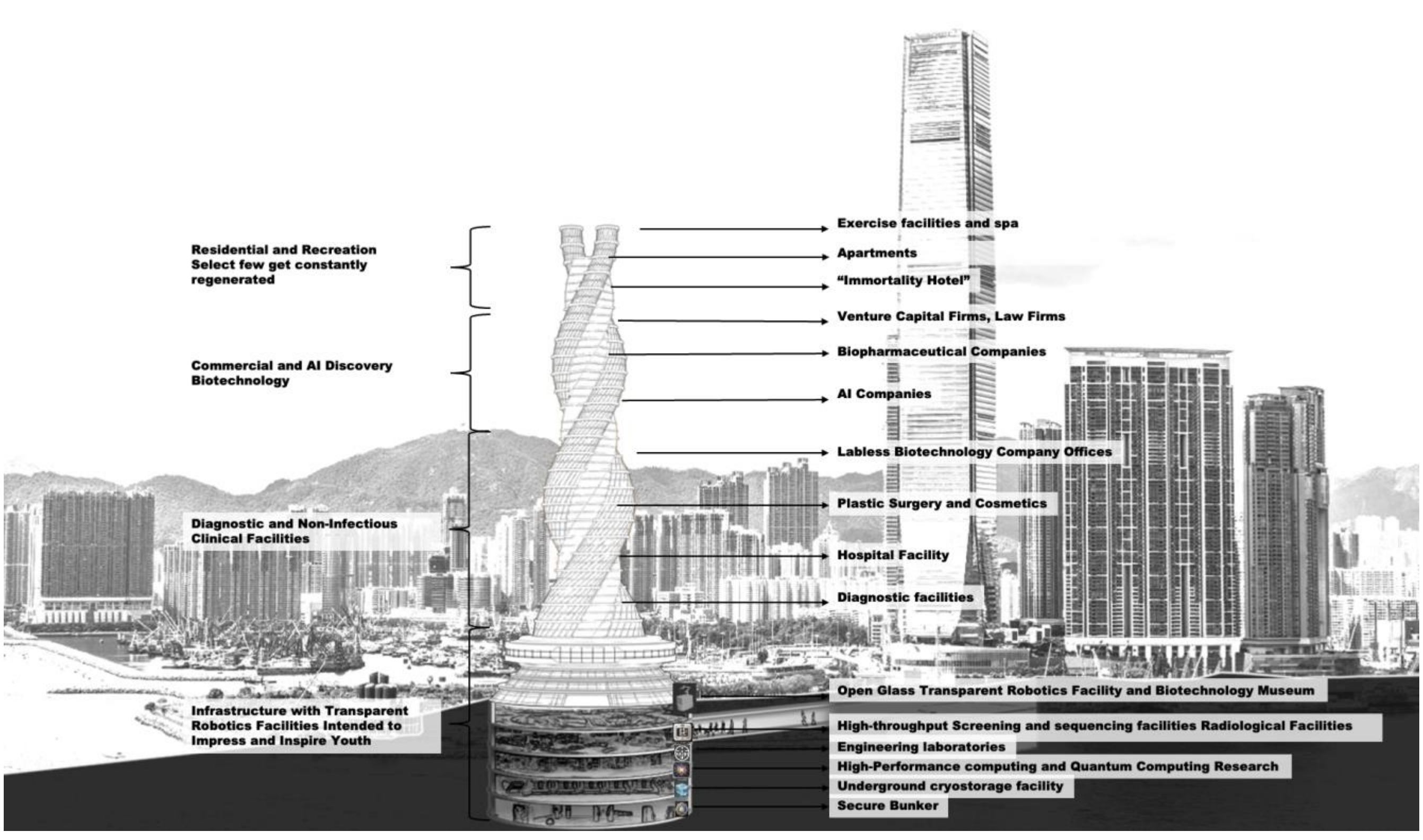

Figure 5: End-to-end vertically-integrated residential biotechnology and biomedical infrastructure imagined in the landmark double-helical structure portrayed in the West Kowloon region of Hong Kong and connected to the existing highly-connected infrastructure. In densely populated cities with scarce high-value real estate the iconic “DNA Towers” may contribute to increased tourism and recognition of efforts in biotechnology by the residents and international community.

# Conclusion

This chapter has outlined a vision for an Integrated Biotechnology Hub that unites real estate, healthcare, research, and artificial intelligence into a seamless ecosystem. By addressing the inefficiencies of current models and leveraging the power of multimodal data collection, advanced AI, and purpose-built infrastructure, the proposed framework has the potential to transform pharmaceutical innovation and longevity research. Instead of fragmented and siloed efforts, this integrated model creates an environment where talent, capital, and health data converge to accelerate discovery, improve patient outcomes, and enhance the overall value of biomedical research. Importantly, this shift reframes real estate not as a passive asset but as an active enabler of scientific progress and societal well-being.

Looking forward, the challenge lies in translating this ambitious model into reality through bold collaboration between governments, real estate developers, biotechnology companies, and research institutions. Emerging examples such as NEOM in Saudi Arabia and TopTown in China provide early testbeds for several core concepts, but global adoption will require trust in data stewardship, dynamic governance, and a shared commitment to ethical innovation. As populations age and healthcare systems face mounting strain, the urgency for reimagining biomedical R&D becomes even more pronounced. The Integrated Hub presents not just an incremental improvement, but a fundamentally new paradigm that could position entire cities as engines of longevity, resilience, and economic growth. By embracing this model, we can turn the places we live and work into engines powering biomedical innovation and longevity.

**Acknowledgements**

We thank Dr. David Gennert for assistance in writing the present chapter.

**Compliance with ethical standards**

**Conflicts of interest**
AZ is employed by Insilico Medicine, a company developing an AI-based end-to-end integrated pipeline for drug discovery and development that is engaged in drug discovery programs for aging, fibrosis and oncology. CYL is employed by Silver Dart Capital and Value Partners Private Equity Limited, and is Vice President of the Hong Kong Biotechnology Organization (HKBIO).